\documentclass[aps,prl,showpacs,twocolumn,superscriptaddress,longbibliography,amsmath,amssymb]{revtex4-2}
\usepackage[pdftex]{graphicx}
\usepackage[pdftex,bookmarks,colorlinks,breaklinks]{hyperref}
\usepackage{amsmath,amssymb}
\usepackage{bbold}
\usepackage{color}
\hypersetup{linkcolor=red,citecolor=blue,filecolor=blue,urlcolor=blue}
\hypersetup{pdftoolbar=true,pdfpagemode=UseNone,pdfstartview=FitH,colorlinks=true,extension=}
\newcommand{\uVNA}{_\mathrm{VNA}}
\newcommand{\uNLCPA}{_\mathrm{NLCPA}}
\newcommand{\uIQ}[1]{_\mathrm{IQ#1}}
\newcommand{\uNC}{_\mathrm{NC}}
\newcommand{\uEP}{_\mathrm{EP}}
\newcommand{\ucal}{_{cal}}
\newcommand{\ueff}{_{eff}}

\begin{document}

\title{Towards a Broad-Band Coherent Perfect Absorption in systems without Scale-Invariance}
\author{Suwun Suwunnarat}
\affiliation{Wave Transport in Complex Systems Lab, Physics Department, Wesleyan University, Middletown CT-06459, USA}
\author{Yaqian Tang}
\affiliation{Wave Transport in Complex Systems Lab, Physics Department, Wesleyan University, Middletown CT-06459, USA}
\author{Mattis Reisner}
\affiliation{Universit\'{e} C\^{o}te d'Azur, CNRS, Institut de Physique de Nice (INPHYNI), 06108 Nice, France, EU}
\author{Fabrice Mortessagne}
\affiliation{Universit\'{e} C\^{o}te d'Azur, CNRS, Institut de Physique de Nice (INPHYNI), 06108 Nice, France, EU}
\author{Ulrich Kuhl}
\thanks{ulrich.kuhl@univ-cotedazur.fr}
\affiliation{Universit\'{e} C\^{o}te d'Azur, CNRS, Institut de Physique de Nice (INPHYNI), 06108 Nice, France, EU}
\author{Tsampikos Kottos}
\thanks{tkottos@wesleyan.edu}
\affiliation{Wave Transport in Complex Systems Lab, Physics Department, Wesleyan University, Middletown CT-06459, USA}

\begin{abstract}
We experimentally and theoretically challenge the concept of coherent perfect absorption (CPA) as a narrow frequency resonant mechanism associated with scattering processes that respect scale-invariance. 
Using a microwave platform, consisting of a lossy nonlinear resonator coupled to two interrogating antennas, we show that a coherent incident excitation can trigger a self-induced perfect absorption once its intensity exceeds a critical value. 
Importantly, a (near) perfect absorption persists for a broad band frequency range around the nonlinear CPA condition. 
Its origin is traced to a quartic behavior that the absorbance spectrum acquires in the proximity of a CPA associated with a new kind of exceptional point degeneracy related to the zeros of the nonlinear scattering operator. 
\end{abstract}

\maketitle

\emph{Introduction -- } 
Perfect absorption of the energy carried by a classical wave is an interdisciplinary research theme spanning areas as diverse as acoustics and mechanical waves \cite{mei12,ma14}, to radio-frequency \cite{sal52}, microwaves \cite{lan08,pha13} and optical wave settings \cite{cai00,pip14,stu16,zho16,bar17}. 
At the core of this activity is the promise that perfect absorption can be beneficial to a variety of applications ranging from stealth technologies \cite{sal52,fan88,vin96}, energy harvesting and photovoltaics \cite{kat16} to sensing \cite{liu10,kra13} and photodetection \cite{kom10,kni11}. 
Along these lines, the quest for low-cost/power all-optical switching and modulation schemes which simultaneously utilize coherent interaction of light beams and absorbing matter for extreme absorption is recently gaining a lot of attention \cite{zha12,fan15,zha16}. 
One such protocol, is the so-called coherent perfect absorption.

Coherent perfect absorption (CPA) is a multichannel waveform shaping protocol which leads to a complete extinction of a monochromatic radiation when it enters a weakly lossy cavity \cite{cho10,wan11b,bar17}. 
Although the scheme has been initially proposed in the framework of classical optics \cite{cho10,wan11b}, as the time-reversed process of a laser, it turns out that its implementation does not require time-reversal symmetry \cite{fyo17,li17,arXche20}. 
It rather solely relies on wave interference effects that entrap the incident radiation inside the lossy cavity, leading to its complete absorption. 
Subsequent studies nicely demonstrated the CPA implementation, beyond the original platform of optics \cite{wan11b,zha12,pu12,yoo12,bru13,rog15,won16,stu16,zha16}, spanning all areas of classical wave physics ranging from microwave \cite{arXche20,pic19,sun14} and RF \cite{sch12b}, to acoustics \cite{wei14b,rom16}. 
In all above mentioned cases, the CPA protocol demonstrated a narrow, resonant-based, (perfect) absorption with very sharp characteristics around the frequency of perfect absorbance. 
Obviously, addressing this ``deficiency'' will open-up a whole new range of possibilities for the CPA scheme including solar photovoltaic or stealth applications. 

At the same time, most of the CPA studies and its implementations have been performed under the assumption of scale invariance, i.e., the underlying wave systems were linear. 
The absence of implementation of CPA protocols in wave systems that lack scale-invariance comes as a surprise, specifically since non-linear mechanisms are abundant in nature and they offer additional degrees of freedom for light manipulation. 
Only recently, some researchers \cite{muel18,red13,ach16} have put forward the question of the applicability of a CPA scheme in cases where scale invariance is violated. 
The reasons for this lack of effort to identify nonlinear CPA (NLCPA) protocols is two-fold: 
From the theoretical side, one needs to develop new computational schemes since the well established (linear) scattering formalism is not any more applicable. 
Moreover, in the presence of non-linear interactions one needs to control, not only the relative phases and amplitudes of the incident waves, but also their absolute magnitude. 
From the experimental side, one might question the viability of such protocols due to bi-stabilities and other nonlinearity-driven phenomena which might destroy the delicate interferences between various propagating waves, or result in modulation instabilities making the NLCPA concept unrealistic.
It is therefore imperative to test the implementation of a non-linear CPA protocol (if at all realizable) under experimental conditions. 

In this paper we introduce an electromagnetic platform, consisting of microwave resonators, where a NLCPA with broad-band characteristics can be investigated theoretically and implemented experimentally. 
The absence of scale invariance and the dependence of the scattering process from the absolute magnitude of the incident wave amplitudes result to a {\it self-induced CPA} with a variety of photonic applications in areas like signal processing, non-linear interferometry and sensing. 
In the case of perfect coupling of the resonators with the interrogating antennas, the system supports a new type of NLCPA modes which demonstrate a square-root frequency degeneracy in the neighborhood of a critical magnitude of the incident wave amplitudes-- remnant of an exceptional point degeneracy occurring in linear non-hermitian systems. 
These steady-state non-linear solutions of the wave operator, with incoming boundary conditions, are responsible for a {\it broad-band (near-) perfect absorption}, which challenges the intuitive understanding of linear CPA as resonant (narrow-band) absorbing protocols.

\begin{figure}
	\centerline{
	\includegraphics[width=.8\columnwidth]{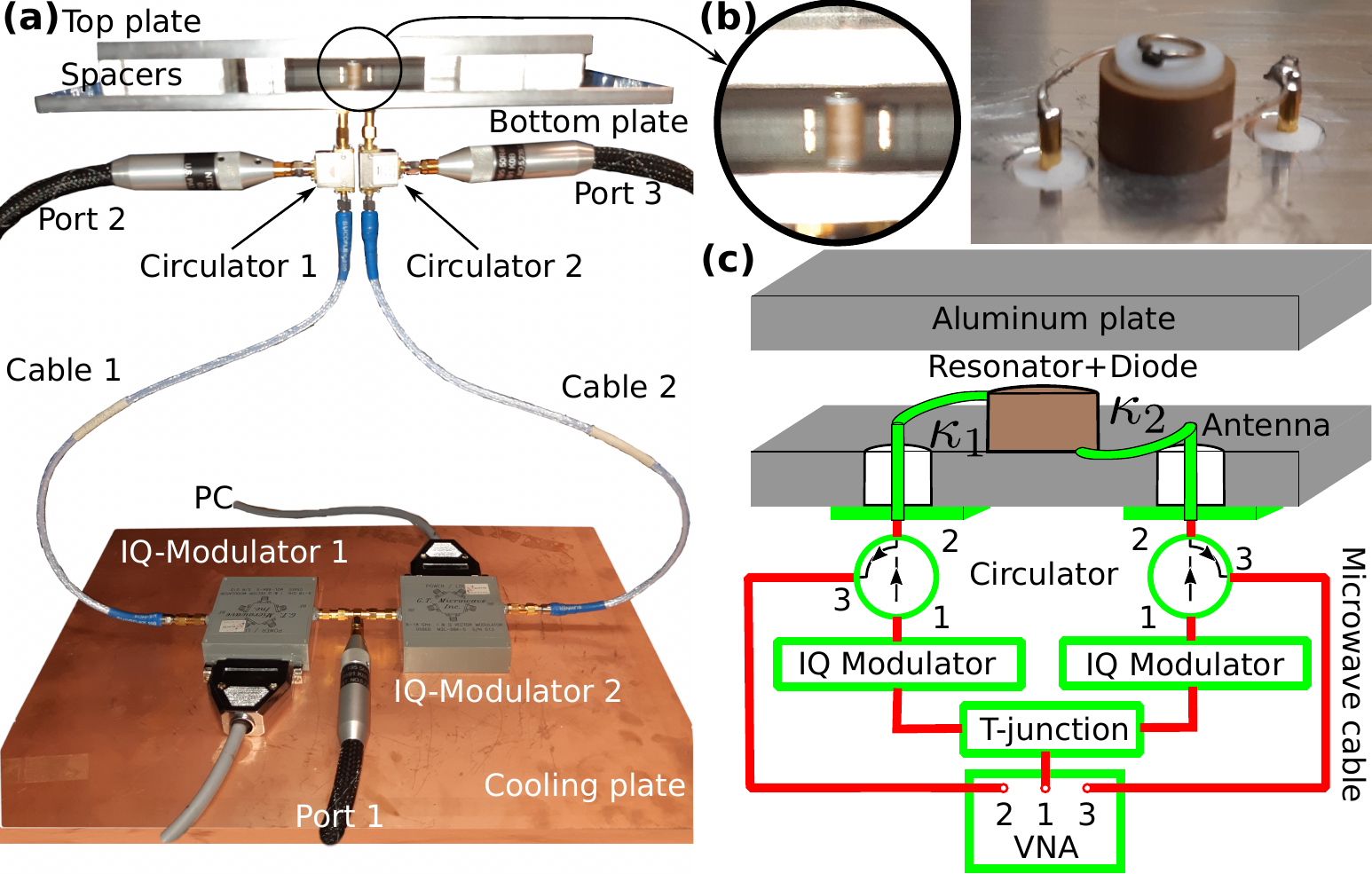}
	}
	\caption{ \label{fig:fig1_setup}
		The experimental setup. 
		a) Photograph of the experimental system including the connections. 
		b) Zoom showing the experimental non-linear system consisting of a cylindrical dielectric resonator where a short-circuited diode is fixed above via a 1\,mm high Teflon spacer with the two transmitting/receiving antennas. 
		c) Sketch detailing the setup.
	}
\end{figure}

\emph{Experimental Setup -- }
The experimental setup [see Fig.~\ref{fig:fig1_setup}a] consists of a dielectric resonator coupled to a short-circuited diode [Fig.~\ref{fig:fig1_setup}b] sandwiched between two aluminum plates. 
This hybrid system demonstrates a non-linear response (see Supplement) \cite{rei20,jeo20}. 
The resonator is excited via two kink antennas curved around it. 
The antennas excite the first TE-resonance mode of the resonator (around 6.7\,GHz), where the magnetic field $\vec{B}$ has only a $z$-component and the electric field lies parallel to the aluminum plates. 
The magnetic field couples to the short-circuited diode thus inducing a non-linear behavior of the system. 
The incident and reflected waves of the system are separated by circulators connected to the source cable, the antenna, and the measuring port of the vector network analyzer (VNA). 
The excitation, with power $P\uVNA$, is injected from port 1 of the VNA and it is splitted equally by a T-junction going through the IQ-modulators into the circulators. 
The IQ-modulators allow to vary the power and phase difference between the excitation lines. 
Thus the measured complex transmission amplitudes from port 1 to port 2 $S_{21}$ and to port 3 $S_{31}$ of the VNA give access to the reflected power of the system $R=|S_{21}|^2+|S_{31}|^2$. 
The total absorbance is then given by $A=1-R=1-|S_{21}|^2 - |S_{31}|^2$ where the scattering matrix elements $|S_{21}|^2$ and  $|S_{31}|^2$ have been normalized taking into account the absorbance due to the IQ-modulator, the cables, and the circulators (for details see supplementary material section 2). 

The power of the incident signals $P\uIQ1$ and $P\uIQ2$, injected from each of the two antennas are controlled by the power injected from the VNA $P\uVNA$ (maximal accessible $P\uVNA=10$\,dBm), and by a differential absorption ${\cal P}\uIQ1$ and ${\cal P}\uIQ2$ associated to each of the two IQ-modulators adjoint to the two antennas. 
The same IQ-modulators can also tune the relative phases $\phi\uIQ1$ and $\phi\uIQ2$ of each of the two incident waves. 
Finally, the couplings between the resonator and the two antennas have been treated as free parameters and they have been adjusted by curving the horizontal part of the kink antennas (see Fig.~\ref{fig:fig1_setup}) and/or by appropriately positioning the dielectric resonator in their proximity.

\emph{Theoretical Model -- } 
The transport characteristics of a system of coupled resonators is modeled using a time-independent coupled mode theory (CMT) 
\begin{equation} \label{eq:timeindep}
\omega\psi_n= H_{n,n-1} \psi_{n-1}+H_{n,n+1} \psi_{n+1}+H_{n,n} \psi_n\,,
\end{equation}
where $\omega$ is the frequency of the incident monochromatic wave, $\psi_n$ is the scattering magnetic field amplitude represented in the TE-modes of the individual resonator (Wannier basis) localized at the $n$-th resonator \cite{bel13b}, $H_{n,n+1}=H_{n+1,n}^*$ describes the coupling between the $n$-th and $n+1$-th resonators and $H_{n,n}=\epsilon_n$ is the resonant frequency of the $n-$th resonator. 
For the specific case of Fig.~\ref{fig:fig1_setup}a) the scattering system consists of only one resonator at $n=0$. 
The two semi-infinite chains of coupled resonators with $n\neq0$ have the same resonant frequency $\epsilon_n=\epsilon\approx 6.7$\,GHz and model the left and right antennas (leads). 
They are coupled with one-another via a (real) coupling constant $H_{n,n+1}=\nu$ giving rise to a dispersion relation $\omega=\epsilon+2\nu \cos(k)$ ($k\in [0,2\pi]$ is the wave-vector of a propagating wave). 
The coupling strength between the nonlinear resonator ($n=0$) with the left ($n<0$) and to the right ($n>0$) leads are $H_{-1,0}=\kappa_{1}=|\kappa_1|e^{-i\phi_1}$ and $H_{0,1}=\kappa_2=|\kappa_2|e^{-i\phi_2}$, respectively. 
The nonlinear resonator at site $n=0$, has a resonant frequency $\epsilon_0=\epsilon+\Omega(|\psi_0|^2)$. 
The complex function $\Omega(|\psi_0|^2)=\Omega_r(|\psi_0|^2)+i\Omega_i(|\psi_0|^2)$ depends on the local field intensity $|\psi_0|^2$. 
Measurements of one-sided transmission measurements indicated that the best fit with the predictions of the model of Eq.~(\ref{eq:timeindep}) are achieved when 
\begin{equation} \label{nlres}
\Omega(|\psi_0|^2) \approx \beta_0+\beta_1 |\psi_0|^2 
\end{equation}
where $\beta_0=(67.74+3i)$\,MHz, $\beta_1=(-0.141+0.41i)$\,MHz/mW while $\nu=0.1$\,GHz.

The most general solution of Eq.~(\ref{eq:timeindep}) in the left and right leads, can be written as
\begin{equation} \label{eq:generalwave}
\psi_n=\left\{
\begin{aligned}
I_1 e^{ikn}+ R_1 e^{-ikn}&      & {n<0}\\
I_2 e^{-ikn}+R_2 e^{ikn}&      & {n>0}
\end{aligned}
\right.\,,
\end{equation}
where $I_{1,2}$ are given by the scattering boundary conditions and represent the amplitudes of a left/right incident waves. 
Substituting the above expressions in Eq.~(\ref{eq:timeindep}) for $n=0$, we get:
\begin{eqnarray} \nonumber 
\left[\omega-\Omega\left(\left|\psi_0\right|^2\right)\right]\psi_0
&=&\ \ \kappa_1^*\left(I_1 e^{-ik} +R_1 e^{ik}\right)\\
&&+\kappa_2\left(I_2 e^{-ik}+R_2 e^{ik}\right)
\label{eq:general}
\end{eqnarray}
An additional relation between $R_{1,2}$ and $I_{1,2}$ is derived by substituting Eq.~(\ref{eq:generalwave}) back in Eq.~(\ref{eq:timeindep}) for $n=\pm1$. 
We have that
\begin{equation} \label{eq:psi0}
\psi_0=\frac{\nu}{\kappa_1} \left(I_1+ R_1\right)=\frac{\nu}{\kappa_2^*} \left(I_2 + R_2\right)\,,
\end{equation}
which allows us to express Eq.~(\ref{eq:general}) in terms of the monochromatic frequency $\omega$ and the amplitudes $I_{1,2}$ and $R_{1,2}$ of the counterpropagating waves in each of the two leads.

A CPA protocol inhibits all outgoing waves, i.e., $R_1=0=R_2$.
Imposing these constrains in Eq.~(\ref{eq:psi0}) allow us to express $\psi_0$ as $\psi_0=\frac{\nu}{\kappa_1}I_1=\frac{\nu}{\kappa_2^*}I_2$.
This relation indicates that the field $\psi_0$ (and therefore the nonlinear losses $\Omega(|\psi_0|^2)$) is controlled by the couplings and the incident amplitudes $I_{1,2}$ of the incoming waves while it remains unaffected by $\omega$. 
A re-arrangement of the above relation into an expression for the relative amplitudes leads us to the conclusion that a potential CPA occurs only if the condition $I_2=\frac{\kappa_2^*} {\kappa_1}{I_1}$ is satisfied. 
For symmetric couplings with $|\kappa_1|=|\kappa_2|=\kappa_0$ we have $I_2={I_1}e^{i\phi}$ where the relative phase is $\phi=(\phi_1+\phi_2$). 
When we substitute these expressions in Eq.~(\ref{eq:general}), together with the CPA constraint $R_1=R_2=0$, we arrive at a transcendental equation with respect to $\omega$, whose (complex) roots are functions of the incident field intensity $|I_1|^2$ and can be associated with a NLCPA. 
In fact, {\it only their real value subset} (if any!) of these $\omega$-roots are physically admissible CPA solutions as they are the only ones that satisfy the incoming boundary conditions (i.e.~propagating waves). 
We point out that as opposed to the linear CPA, here the $I_1$ (or $I_2$) is treated as a free parameter that can enforce a NLCPA. 

%%%%%%%%%%%%%%%%%%%%%%%%%%%%%%%%%%%%%%%%%
\emph{Absorbance -- }
To confirm the efficiency of the NLCPA protocol, we first analyze the total absorbance $A(\omega;I_1,I_2)$:
\begin{equation} \label{eq:theta}
A(\omega;I_1,I_2)=1-\frac{|R_1|^2+|R_2|^2}{|I_1|^2+|I_2|^2} \,,
\end{equation}
where $\omega\in {\cal R}$ are the real frequencies of an incident monochromatic wave with left/right amplitudes $I_1$ and $I_2$, respectively. 
For one-side incident waveforms (e.g.~$I_2=0$) the absorbance gets the expected form $A=1-|r|^2-|t|^2$ where $|r|^2,|t|^2$ are the reflectance and transmittance respectively. 
In the multichannel case, the evaluation of $A(\omega;I_1,I_2)$ requires the knowledge of $R_1, R_2$ which, for a given set $\omega;I_1,I_2$, can be calculated via Eqs~(\ref{eq:general},\ref{eq:psi0}). 
Note that the absorbance $A$ in Eq.~(\ref{eq:theta}) acquires the maximum value $A=1$ whenever $R_1=R_2=0$, i.e., when we have a NLCPA condition. 

In order to minimize the available parameter space, we strived to achieve symmetric coupling amplitude configurations, i.e., $|\kappa_1|=|\kappa_2|=\kappa_0$. 
The latter has been guaranteed via weak power measurements (linear scattering regime), when the reflected signals at each antenna individually was measured to be the same. 
We have further simplified our interrogation scheme by setting ${\cal P}\uIQ1=1$\,dB and $\phi\uIQ1=0$ while varying the amplitude and phase ${\cal P}\uIQ2, \phi\uIQ2$ (via the second IQ-modulator) of the injected signal through the second antenna together with the total absolute power $P\uVNA$ controlled by the VNA. 
Finally, we have scanned the residual parameter space and measured for each of the varying parameters the transmissions $S_{21}$ and $S_{31}$ from which we extracted the total reflected power and absorbance $A$. 

\begin{figure}
	\centerline{\includegraphics[width=.9\linewidth]{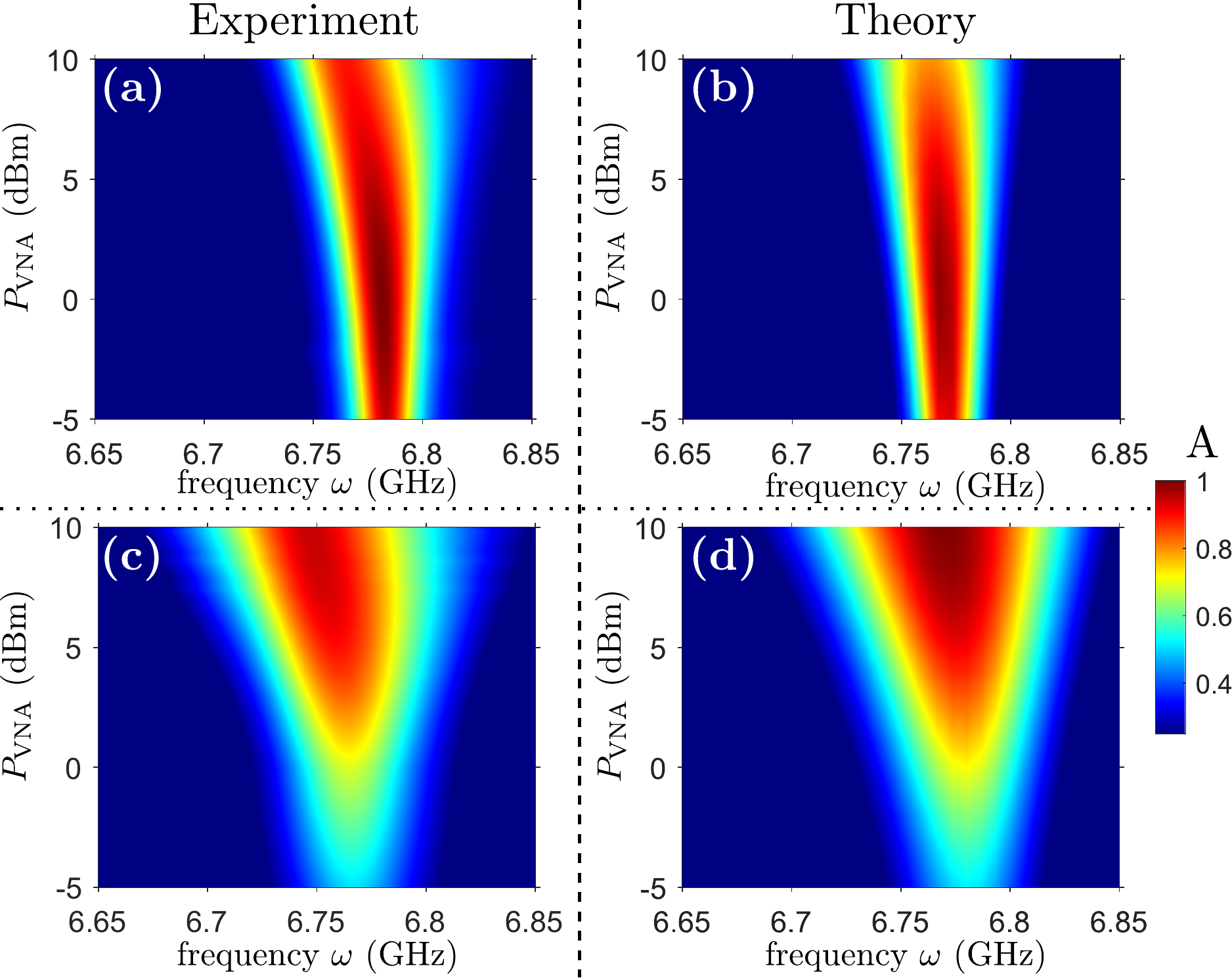}}
	\caption{ \label{fig:fig2_exp_abs}
	Absorbance $A$ as a function of frequency $\omega$ and the injected power $P\uVNA$. 
	Left column (A,C) shows our measurements and the right column (B,D) shows the results of our modeling Eq.~(\ref{eq:timeindep}) with $\Omega$ given by Eq.~(\ref{nlres}). 
	(A,B) The two antennas are weakly coupled ($\kappa_0=0.23\nu$). 
	The experimental maximal absorbance is $A=99.99$\% occuring at $P\uNLCPA=0$\,dBm, $\omega\uNLCPA\approx6.782$\,GHz. 
	The theoretical NLPCA's gives $A=1$, at $P\uNLCPA=-0.486\,\mathrm{dBm}$, $\omega\uNLCPA =6.769\,\mathrm{GHz}$. 
	(C,D) The two antennas are moderately coupled ($\kappa_0=0.4\nu$). 
	The experimental maximum absorbance is $A\approx 95$\% occuring at $P\uNLCPA=9.8$\,dBm, $\omega\uNLCPA=6.747$\,GHz. 
	The theoretical NLPCA with $A=1$, occurs at $P\uNLCPA=10.23$dBm, $\omega\uNLCPA=6.77$\,GHz. 
	}
\end{figure}

From the previous discussion, we expect that when $\sqrt{{\cal P}\uIQ1/{\cal P}\uIQ2}=\left|\kappa_1/\kappa_2\right|= 1$ the system might support a NLCPA at some critical incident power $P\uVNA$. 
Of course, a necessary condition is that the extracted NLCPA frequency $\omega\uNLCPA$ is real. 
To this end, we proceed with the analysis of the absorbance measurements for two settings associated with a weak and moderate coupling constants $\kappa_0$. 
In the former case, we find a set of parameters $P\uVNA= 0$\,dBm, ${\cal P}\uIQ1=5.0$\,dB, ${\cal P}\uIQ2=5.7$\,dB, and $\phi\uIQ2=92^\circ$ for which $A\geq 99.99\%$ at  $\omega_\mathrm{NC}=6.782$\,GHz. 
The slight difference between left and right incident wave powers is attributed to the fact that $\left|\kappa_1/\kappa_2\right|\approx 1$. 
The same extreme absorption is found for moderate coupling constants, for which $A\approx 95\%$. 
The corresponding NLCPA parameters are $\omega\uNC=6.747\,\mathrm{GHz}, P\uVNA= 9.8\,\mathrm{dBm}, {\cal P}\uIQ1=5.0\,\mathrm{dB}, \phi\uIQ1=0^\circ, {\cal P}\uIQ2=5.0\,\mathrm{dB}, \phi\uIQ2=83.5^\circ$. 
In both cases a shift of the maximal absorbance as a function of the incident power is observed showing that the non-linearity induces also a slight frequency shift. 
In Fig.~\ref{fig:fig2_exp_abs} (left column) we report the measured absorbance as a function of frequency $\omega$ and incident power $P\uVNA$. 
In these measurements the relative phases of the incident waves were kept fixed, given by $\phi\uIQ2=92^\circ$ and $\phi\uIQ2=83.5^\circ$ for weak and moderate couplings respectively. 
We have checked that variations of the relative phase and amplitudes of the incident waves result in a rapid deterioration of the absorbance -- thus underlying the delicate interferometric process between the two left and right monochromatic waves (see Supplement).  

Next, we turn to the theoretical analysis of $A(\omega; I_1,I_2)$. 
In our modeling, $\kappa_1=\kappa_0$ and $\kappa_2=-i\kappa_0$ with $\kappa_0=0.23$\,GHz for the weak and $\kappa_0=0.04$\,GHz for the moderate couplings, see Fig.~\ref{fig:fig2_exp_abs} (right column). 
We considered that $I_1=I$ and $I_2=e^{i\phi}I_1$ with relative phase $\phi=\pi/2$. 
Using Eqs~(\ref{eq:general},\ref{eq:psi0}) we extracted numerically the corresponding $R_1,R_2$ and evaluated the absorption $A(\omega; I)$ using Eq.~(\ref{eq:theta}). 
We have achieved perfect absorption $A=1$ at $(I\uNLCPA,\omega\uNLCPA)$ corresponding to $(-0.48\,\mathrm{dBm}, 6.769\,\mathrm{GHz})$ and $(10.23\,\mathrm{dBm}, 6.770\,\mathrm{GHz})$ for weak and moderate coupling values $\kappa_0$ respectively. 
These results are in quantitative agreement with the experiment. 
A surprising feature of our calculations, which reproduces the behavior of the measured absorbance $A(\omega)$, is the broadening of the frequency range over which large absorption values are achieved as $\kappa_0$ approaches perfect coupling $\kappa_0=\nu$. 
This domain appears to be in the vicinity of the NLCPA and tends to occur for larger $I\uNLCPA$ values as $\kappa_0$ increase. 
For example, for $\kappa_0=0.04$\,GHz, one has that $A(\omega\in[\omega\uNLCPA \pm 21\,\mathrm{MHz}]>80\%$ occurring at incident powers around $P\uVNA\approx10\,\mathrm{dBm}$ which is the maximum power of our VNA. 

%%%%%%%%%%%%%%%%%%%%%%%%%%%%%%%%%%%%%%
\emph{Absorption broadening due to EP degeneracies of NLCPAs --}
To understand the origin of the broad-band high absorptivity we analyze the parametric evolution of the NLCPA frequencies versus the incident power $I$. 
By imposing the CPA conditions $R_1=R_2=0$ and combining Eqs~(\ref{eq:general},\ref{eq:psi0}) we arrive to the transcendental equation for $\omega\uNLCPA$ 
\begin{equation} \label{eq:1scatter_w}
  f(\omega)=\Omega\left(|\frac{\nu}{\kappa_1}I_1|^2\right)+\frac{\kappa^2}{\nu} e^{-ik}-\omega=0,
\end{equation} 
where $\kappa^2\equiv\left|\kappa_1\right|^2+ \left|\kappa_2\right|^2$ and $\omega(k)=2\nu\cos(k)$. 
We re-iterate that physically acceptable NLCPA's correspond to the case where the $\omega\uNLCPA$ roots of eq.~(\ref{eq:1scatter_w}) are real, corresponding to propagating waves (see Supplement). 

\begin{figure}
	\centerline{\includegraphics[width=1\linewidth]{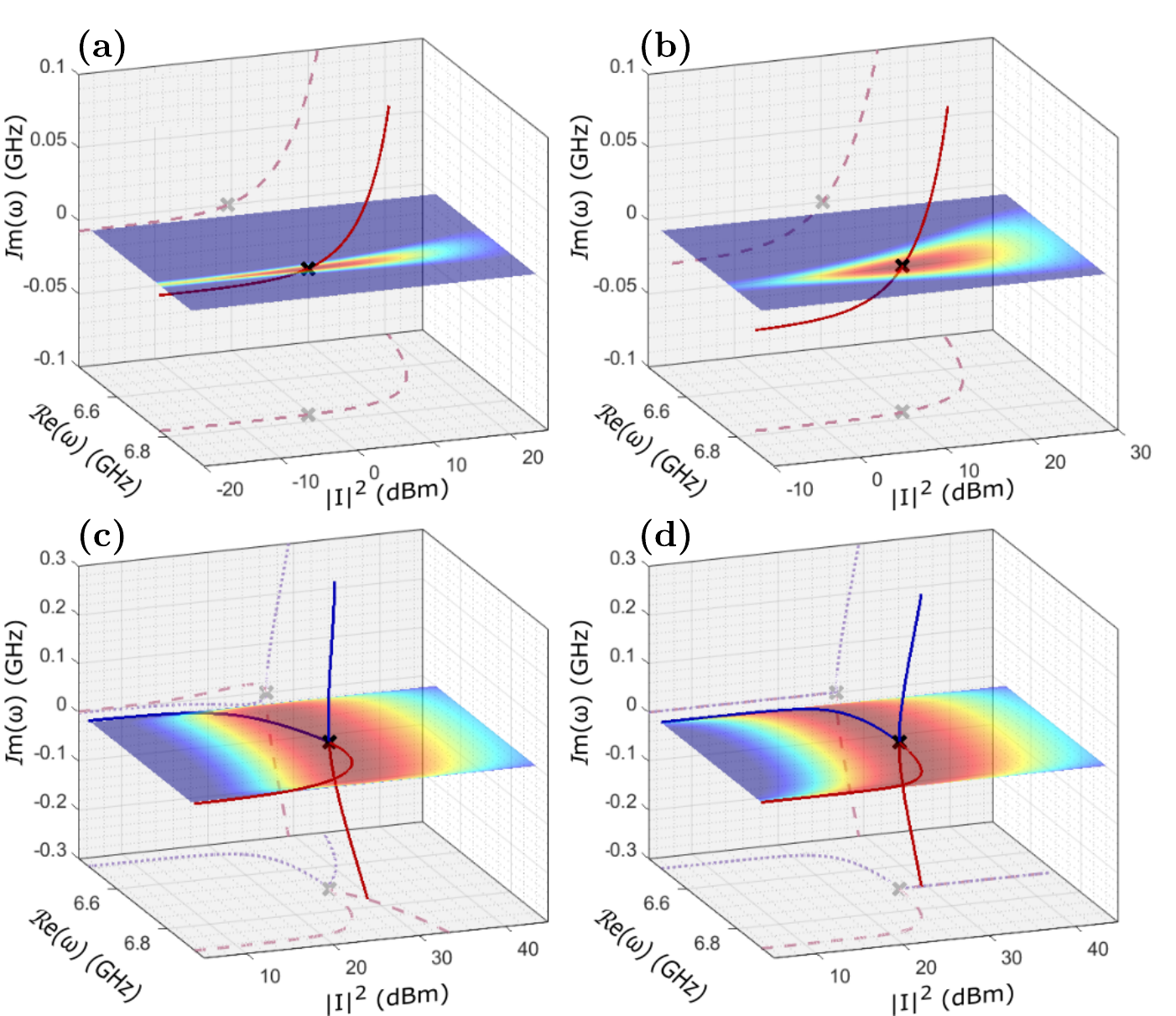}}
	\caption{ \label{fig:fig3_Zeros_NLCPA}
	Trajectories (red and blue solid lines) of the complex zero eigenvalues $\omega$ of the scattering matrix as a function of input power. 
	The dashed and dotted lines are projections of these trajectories on the corresponding plane. 
	The nonlinearity $\Omega=\beta_0 + \beta_1 |\psi_0|^2$ with $\beta_0 = (67.74 + 3.00i) \times 10^{-3}$\,GHz $\beta_1 = (-0.141 + 0.410i) \times 10^{-3}$\,GHz/mW. 
	(A) For two weakly coupled antennas of $\kappa/\nu=0.23$. 
	(B) For two intermediately coupled antennas of $\kappa/\nu=0.40$. 
	(C) For two perfectly coupled antennas $\kappa/\nu=1$. 
	(D) As bottom left but the real part of the nonlinearity is set zero, i.e., no real frequency shift. 
	The absorbance $A$ is reported as a density plot versus the frequency $\omega$ and the power $P\uVNA$ injected by the VNA.
	}
\end{figure}

We consider the specific example of our system where the non-linear resonator takes the form of Eq.~(\ref{nlres}). 
The parametric evolution of the (complex) roots $\omega\uNLCPA$ of Eq.~(\ref{eq:1scatter_w}) vs.~the intensity of the incident wave $I=I_1=I_2$ are shown in Fig.~\ref{fig:fig3_Zeros_NLCPA} (top) for a weak ($\kappa_0/\nu=0.23$) and moderate ($\kappa_0/\nu=0.4$) coupling constants. 
In the frequency range $\omega\in [6.5,6.9]$\,GHz (pass band of the leads), the Eq.~(\ref{eq:1scatter_w}) have only one complex root $\omega$ which crosses the real plane at the incident intensity $I\uNLCPA$ resulting to maximum absorbance $A=1$ (see density plots). 
This is a direct confirmation that the perfect absorption that we have found in our experiment is indeed associated with a NLCPA condition. 
From Fig.~\ref{fig:fig3_Zeros_NLCPA} (top) we see that higher (experimentally inaccessible) intensities of the incident waves suppress the absorbance as they lead to an enhanced impedance mismatch of the resonator. 
The redshift of the maximum absorbance is associated with the real part of the non-linear component of $\Omega (|\psi_0|^2)$. 
Another important conclusion of our analysis is that the critical coupling regime ($\kappa_0=\nu$), enforces an $\omega-I$ parameter domain with high-absorbances. 
This result has been already demonstrated in Fig.~\ref{fig:fig2_exp_abs} but now it is more prominent since we are able to analyze incident powers above $10$dBm (power limit of our VNA).

The situation is more challenging when we consider perfect coupling, i.e., $\kappa_0/\nu=1$. 
In this case, Eq.~(\ref{eq:1scatter_w}) has two complex roots within the propagating band, see Fig.~\ref{fig:fig3_Zeros_NLCPA} (bottom-left). 
One of them (red trajectory) approaches the real plane from above, while the other one (blue trajectory) crosses the real plane from below as the intensity of the incident wave increases. 
At the crossing point (marked with a black cross) the two roots degenerate forming a new type of {\it self-induced exceptional point (EP) degeneracy of CPAs}. 
Its formation demonstrates all the characteristics of an exceptional point (EP) singularity, known from the physics of linear non-Hermitian operators with the most prominent being a square-root singularity. 
Since the EP occurs at the real $\omega-$plane it constitutes a physically allowable CPA with absorbance $A=1$. 
The latter, together with the fact that the two coalescing complex $\omega\uNLCPA$``trajectories'' have small imaginary part (see dashed lines projected in the imaginary $\omega-$plane), results to the appearance of a (near-)perfect (i.e. $A \geq 95\%$) absorption for a broad frequency range covering approximately 90\% of the allowed propagation band of the leads. 
We stress that the broad-band absorption is counter-intuitive and challenges the understanding of CPA as a resonant phenomenon. 
It turns out that the broad band high-absorptivity is quite robust, forgiving small variations of the intensity $|I|^2$ with respect to the critical value. 
  
To understand better the appearance of a broad-band high absorbance, we simplify further our critical coupling set-up by designing the non-linear diode characteristics in order to enforce the two complex-zero trajectories to fall onto the real plane. 
This is shown in Fig.~\ref{fig:fig3_Zeros_NLCPA} (lower right) where the nonlinear parameters in Eq.~(\ref{nlres}) are taken to be $\beta_0 = 3.00i \times 10^{-3}$\,GHz and $\beta_1 = 0.410i \times 10^{-3}$\,GHz/mW. 
In this case, the square-root degeneracy of the NLCPAs $\omega\uNLCPA -\omega\uEP\sim \sqrt{|I\uEP|^2-|I|^2}$ occurs on the real-$\omega$ plane with $\omega\uEP\approx 6.7$\,GHz while $|I\uEP|^2\approx 27$dBm as in lower-left of Fig.~\ref{fig:fig3_Zeros_NLCPA}. 
Such behavior leads to abrupt frequency variations in the vicinity of the $I\uEP$, which span a frequency range as large as 90\% of the available spectrum. 
Of course, once we move away from the critical $I\uEP$-value (say at $|I|^2=15$\,dBm), we recover the typical sharp CPA features with respect to detuning $\omega$ which are reflected in an abrupt drop in absorption for frequency detuning $\omega\neq \omega\uNLCPA$. 

We can quantify the broad-band absorption, by analyzing an effective ``linear'' model which describes the steady-state transport characteristics of our non-linear setting Fig.~\ref{fig:fig1_setup} in the proximity of the EP. 
Specifically, the EP scattering field $\psi_0$, determines the local losses $\Omega_0(|\psi_0|^2)$. 
The corresponding scattering matrix is
\begin{equation} \label{lS}
S(\omega)=-{\hat 1}+\frac{2i\sin(k)}{\nu} W^{\dagger}\frac{1}{H_{eff}-\omega}W=-{\hat 1}+\alpha W^{\dagger}W\,,
\end{equation}
where $H_{eff}=\Omega_0+\frac{e^{ik}}{\nu}WW^{\dagger}, W=(\kappa_1,\kappa_2)$ and $\alpha=\frac{1}{\nu^2}\frac{\sqrt{1-\delta\omega^2}}{1+\sqrt{1-\delta\omega^2}}$ [where $\delta\omega=(\omega-\omega\uEP)/(2\nu)$]. 
In the last equality we assumed EP conditions, i.e., $\kappa_1=\kappa_2=\nu$. 
Interestingly, the condition Eq.~(\ref{eq:1scatter_w}) for the existence of CPA's is equivalent to the eigensolutions of the wave operator (described by $H\ueff$) with incoming boundary conditions ($k\rightarrow -k$), i.e., $\det[H\ueff(-k)-\omega(k)]=0$ \cite{arXche20,fyo17}. 
Therefore, we associate the EP-CPAs with the formation of EPs in the spectrum of $H\ueff(-k)$. 
From Eq.~(\ref{lS}) we have calculated the absorption matrix 
\begin{equation} \label{AL}
A={\hat 1}-S^{\dagger}S
%=\frac{2\alpha}{1+\sqrt{1-\delta\omega^2}}W^{\dagger}W
\approx \frac{1}{2\nu^2}\left(1-\frac{1}{16} \delta\omega^4+{\cal O}
(\delta\omega^6)\right)W^{\dagger}W
\end{equation}
indicating a quartic broadening of the absorption spectrum in the neighborhood of $\omega\uEP$ \cite{swe19} ($\frac{1}{2\nu^2}W^{\dagger}W$ has eigenvalues $1$ and zero). 
The latter prediction is nicely reproduced by the results shown in Fig.~\ref{fig:fig3_Zeros_NLCPA}.

%------------------------------------------------------------
%%%%%%%%%%%%%%%%%%%%%%%%%%%%%%%%%%%%%%%%%%%%%%
\emph{Conclusions -- }
We demonstrated the viability of a self-induced coherent perfect absorber by utilizing a simple microwave setup consisting of a dielectric resonator inductively coupled to a non-linear diode. 
In the weak coupling regime, we have observed sharp (resonant-like) absorption up to 99.99\% at the frequency of the NLCPA. 
As the coupling with the interrogating antennas is approaching the critical coupling regime, the frequency range where extreme absorption ($>95\%$) occurred, increases dramatically leading to a broad-band (near-)perfect absorption in the neighborhood of the NLCPA. 
A CMT model describes the experimental findings nicely and shows that this broadening is attributed to the formation of a new type of self-induced exceptional point degeneracies associated with the NLCPA frequencies. 
It will be interesting to investigate more complex scenarios where the EP of the NLCPA frequencies is of higher order resulting (probably) to an even broader (near-)perfect absorption. 
Our results pave the way to new applications of CPA in microwave/RF and optical regimes.

\begin{acknowledgments}
	Y.T., S.S. and T.K. acknowledge partial support from the Office of Naval Research (Grant No. N00014-19-1-2480) and from a Simons Collaboration in MPS and Y.T. and T.K useful discussions with Do Hyeok Jeon.
\end{acknowledgments}

%-----------------------------------------------------------
\begin{acknowledgments}
	Y.T., S.S. and T.K. acknowledge partial support from the Office of Naval Research (Grant No. N00014-19-1-2480) and from a Simons Collaboration in MPS and Y.T. and T.K useful discussions with Do Hyeok Jeon.
\end{acknowledgments}

%%%%%%%%%%%%%%%%%%%%%%%%%%%%% Supplemental Material %%%%%%%%%%%%%%%%%%%%%%%%%%%%%%%%%
\onecolumngrid
\appendix
\newpage

\centerline{{\Large \bf Supplemental Material}}
\section{Antenna coupling adjustments}
\renewcommand\thefigure{S\arabic{figure}}    
\setcounter{figure}{0}
\setcounter{page}{1}
\renewcommand{\thepage}{S\arabic{page}}
\setcounter{page}{1}
\setcounter{equation}{0}
\renewcommand\theequation{S\arabic{equation}}

We have adjusted and measured the couplings $\kappa_{1,2}$ between the two antennas and the nonlinear resonator by performing small incident power measurements $P\uVNA=-30$\,dBm of the complex transmission coefficients $S_{21}$ and $S_{34}$ from each antenna. 
Under these conditions the scattering process is linear. 
For these measurements, the T-junction has been removed (see Fig.~1) and the port 1 of the analyzer was attached to IQ-modulator 1 while the port 4 was attached to the IQ-modulator 2. 
Thus the transmission amplitudes $S_{21}$ and $S_{34}$ are including the effects of the IQ-modulator, of the microwave cables and of the circulators. Both IQ-modulators have been set to provide zero power reduction and zero additional phase shift.

\begin{figure}[h]
	\centerline{
		\includegraphics[width=.32\linewidth]{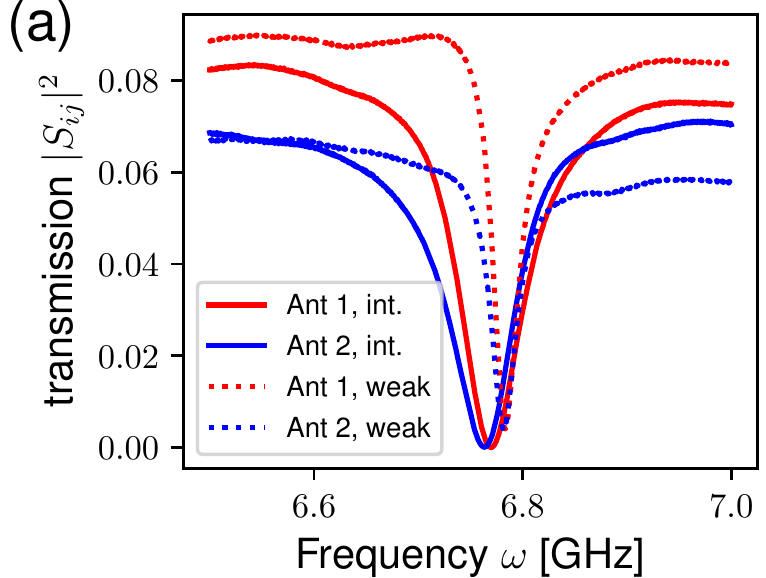}
		\includegraphics[width=.32\linewidth]{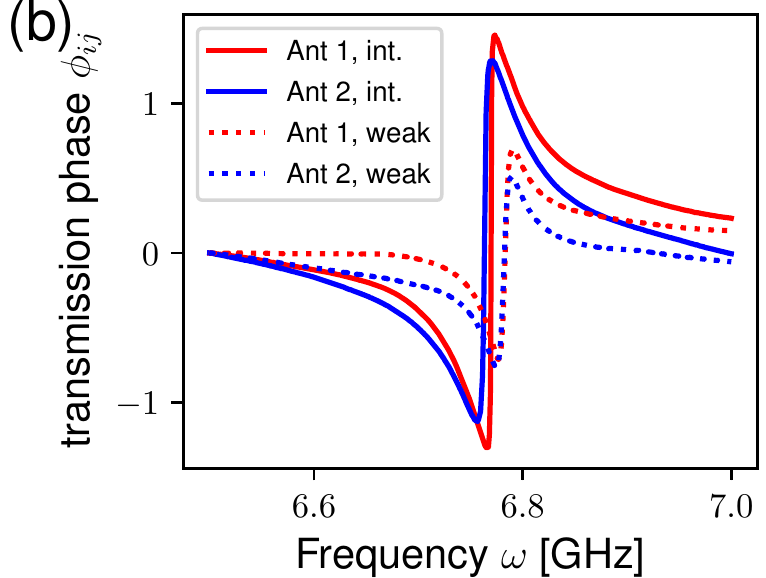}
		\includegraphics[width=.25\linewidth]{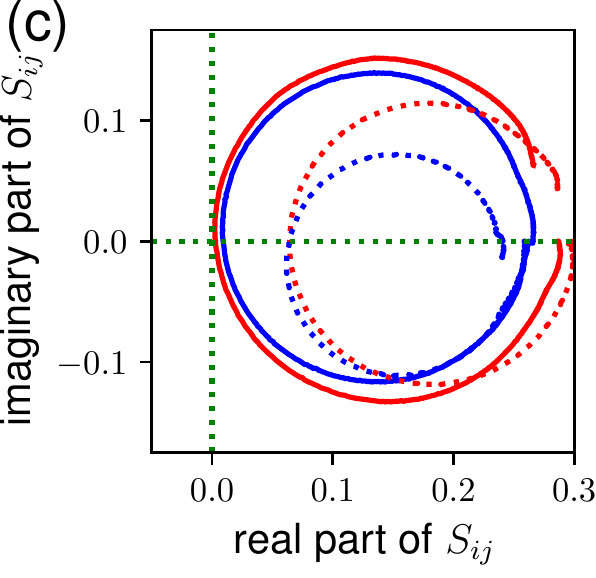}
	}
	\caption{ \label{fig:figs1}
		Transmission measurements to adjust the antenna coupling. Transmission intensities $|S_{ij}|^2$ (a), phase $\phi_{ij}$ (b), and Argand 
		diagram of $S_{ij}$ (c) for the intermediate and weak coupling antenna. Solid lines correspond to the intermediate coupling antenna whereas 
		dotted lines correspond to the weak coupling antennas.
	}
\end{figure}

In a similar manner as typical standard reflection measurements, the transmission coefficients $S_{21}$ and $S_{34}$ are characterizing 
the antennas 1 and 2 respectively respectively. The only difference here is that the baseline is defined by the absorption in the line and 
not at antenna 1. The baseline variation comes from the fact that the antennas are also coupled to the TM$_0$-mode (electrical field in 
the $z$-direction) mainly by their vertical part). As the TM$_0$-mode is approximately freely propagate, it leads to an additional constant 
loss which depends on the individual antenna. 
In Fig.~\ref{fig:figs1} we extract from the $S_{21}$ and $S_{34}$ measurements the 
signal intensity (a), its phase (a), and the Argand diagram ($Re(S)$ as $x$-axis and $Im(S)$ as $y$-axis). The phase has been additionally 
adjusted by removing a global linear phase shift and setting the initial phase to zero. The linear phase shift is fixed to be approximately 
constant at low frequencies for the weakly coupled antenna 2. We used the same linear dependence for all four presented measurements 
which have been obtained by $S_{21}=S^{(exp)}_{21} \exp(-i 24.6 \omega/\mathrm{GHz})$. In the case of weak coupling, our measurements allowed 
us to identify a sharp Lorentz lineshape in the intensity modulus, whereas for stronger couplings the linewidth was enlarged. 

To guarantee similar couplings for antennas 1 and 2 we have given extra care in identifying similar behaviors for $S_{21}$ and $S_{34}$.
The couplings are changed by deforming the horizontal part of each antenna, e.g., by curling more (or less) the antenas around the 
resonator, and by changing the position of the resonator bringing it closer (further away) to the antennas. Figure~\ref{fig:figs1} presents 
the two antenna couplings chosen for the weak and intermediate case discussed in the main text. Their values $\kappa=0.23\nu$ and 
$\kappa=0.4\nu$ have been determined by comparing the frequency dependence of the absorption (Fig.~2)
from the simulations, using coupled mode theory, with the actual experimentally obtained data. 
Of course, one needs to keep in mind that the actual experiment does not incorporates the frequency scale $\nu$ associated with the modeling of the semi-infinite leads and thus the 
$\kappa/\nu$ values can not be determined directly from the experiment.

\section{Transmission calibration}

\begin{figure}[h]
	\centerline{\includegraphics[width=.4\linewidth]{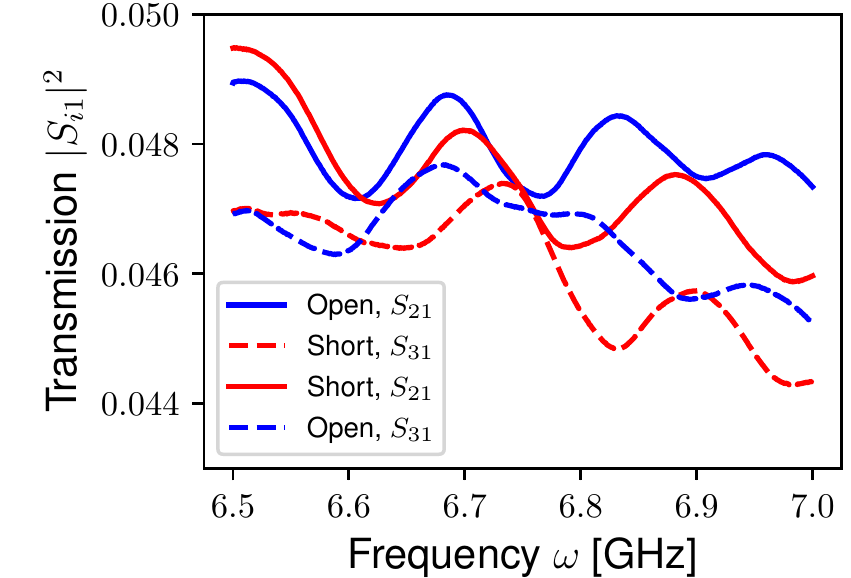}}
	\caption{ \label{fig:figs2}
		Transmission intensities $|S^{(cal)}_{21}|^2$ and $|S^{(cal)}_{31}|^2$ where the nonlinear system has been replaced by open and short terminations at the two antennas, respectively.
	}
\end{figure}

We calibrated the measured transmission by measuring the scattering matrix elements $S^{(cal)}_{21}$ and $S^{(cal)}_{31}$ for the system shown in Fig.~1, where we attached either an open or a short terminator to port 2 of each circulator, thus replacing the connection to the non-linear system.
In Fig.~\ref{fig:figs2} the transmission intensities $|S^{(cal)}_{i1}|^2$ are shown for both open and short termination circuits. 
The incident power $P\uVNA$ has been set to 10\,dBm while all IQ-Modulators are arranged to provide power reduction which is ${\cal 
	P}\uIQ1={\cal P}\uIQ2=0$\,dB together with an additional phase shift $\phi\uIQ1=\phi\uIQ2=0^\circ$.
We have found that the minimal transmission in the interrogating frequency range $\omega\in [6.5,7]$\,GHz is $T\ucal=0.0442$ which incorporates losses from the T-junction, microwave cables, insertion loss of the IQ-modulator, and the losses associated with the double-passing through the circulators. 
Another source of energy loss (which is already included in these measurements) is associated with the open and short terminators. 
We have used the terminators from a VNA calibration kit (Rhode \& Schwarz ZV-Z235). 
The measurements have been repeated for different IQ-modulator values and the measured $S^{(cal)}_{i1}$ spectra 
showed the corresponding power reduction and phase shifts.

The transmission used to evaluate the absorbance has then been normalized by $S_{i1}=S_{i1}^{(exp)}/\sqrt{T\ucal}$. 
This choice of normalization underestimates the experimental absorbance, thus guaranteeing a conservative estimation of our measurements.
\newpage
\section{CPA - parametric dependencies}

The identification of a NLCPA in a multi parameter space is a daunting task. In the main text we have restricted our investigations to a reduced 
parameter space incorporating the frequency and the total injected power. Here we present a number of ``one-dimensional slices'' of absorbance 
versus various ``directions'' of the multi-dimensional parameter space: frequency $\omega$ axis; the total injected power $P\uVNA$; 
the additional power reduction ${\cal P}\uIQ2$ and the phase $\phi\uIQ2$ of the injected wave via antenna 2. 
We also remind that the power reduction for antenna 1 is set to ${\cal P}\uIQ1=5\,\mathrm{dB}$ while the phase is set to be $\phi\uIQ1=0^\circ$.

\begin{figure}[h]
	\centerline{
		\includegraphics[width=.24\linewidth]{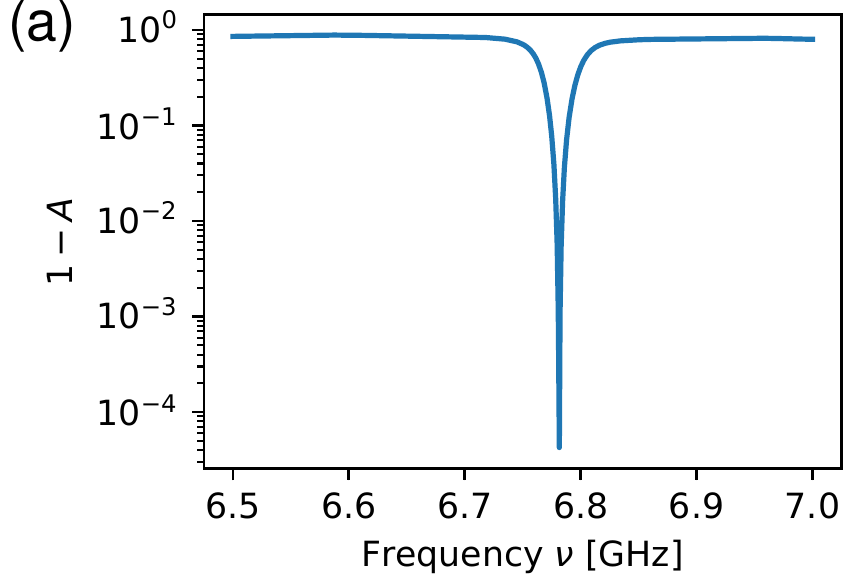}
		\includegraphics[width=.24\linewidth]{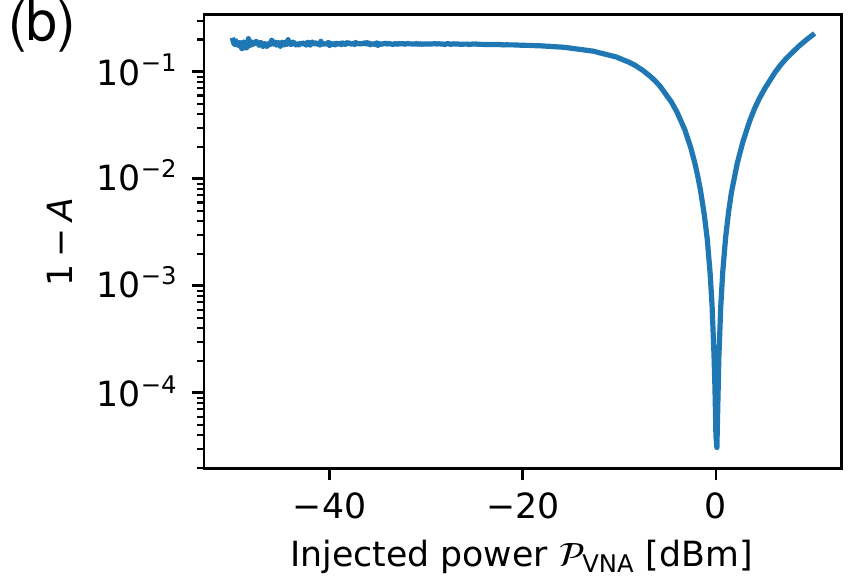}
		\includegraphics[width=.24\linewidth]{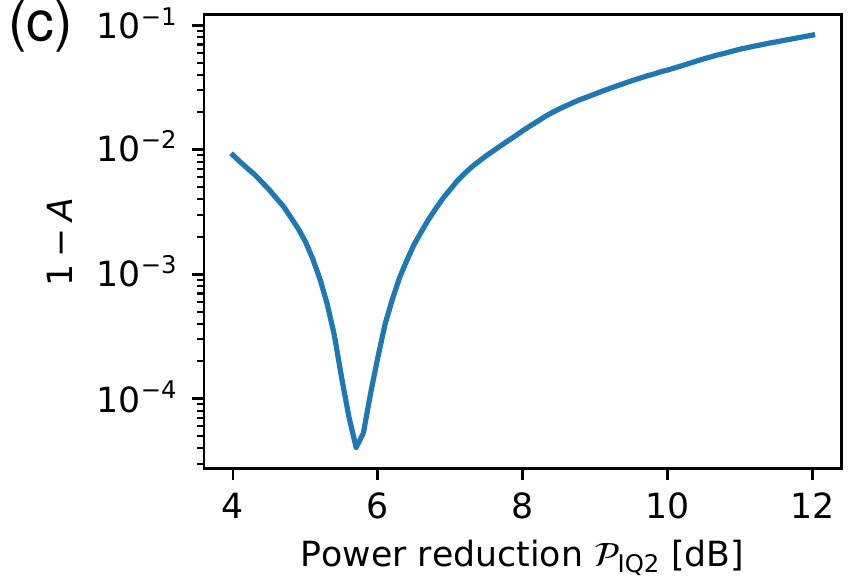}
		\includegraphics[width=.24\linewidth]{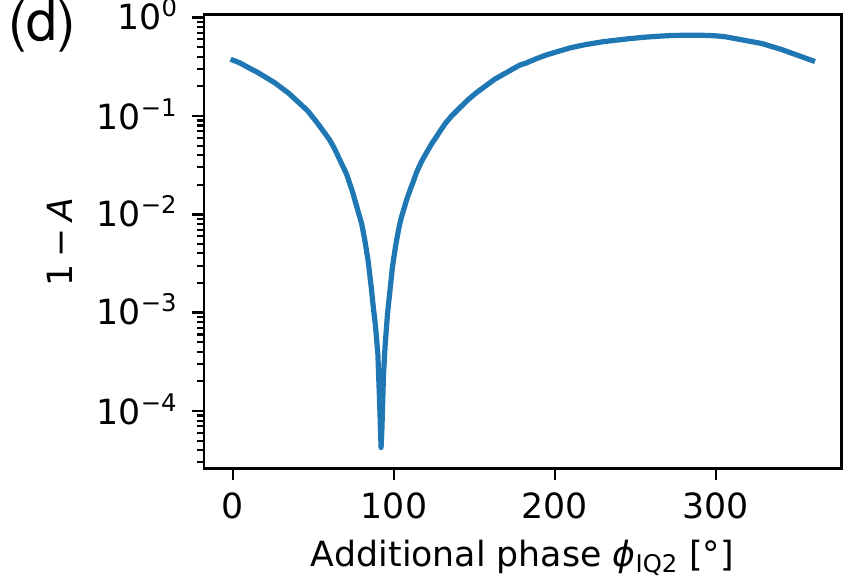}
	}
	\caption{ \label{fig:figs3}
		Deviation from the perfect absorption $1-A$ for the weakly coupled antennas ($\kappa_0=0.23\nu$) as function of the different experimental parameters frequency $\omega$ (a), injected power $P\uVNA$ by the VNA (b), IQ-modulator intensity reduction ${\cal P}\uIQ2$ (c) and phase $\phi\uIQ2$ (d). 	
		The other parameters have been fixed to the value where the maximal absorbance (CPA) has been measured, i.e., 
		$P\uNLCPA=0$\,dBm, $\omega\uNLCPA\approx 6.782$\,GHz, 
		${\cal P}\uIQ1=5$\,dB, $\phi\uIQ1=0^\circ$, 
		${\cal P}\uIQ2=5.7$\,dB, and $\phi\uIQ2=92^\circ$
		, respectively.
	}
\end{figure}

\begin{figure}[h]
	\centerline{
		\includegraphics[width=.24\linewidth]{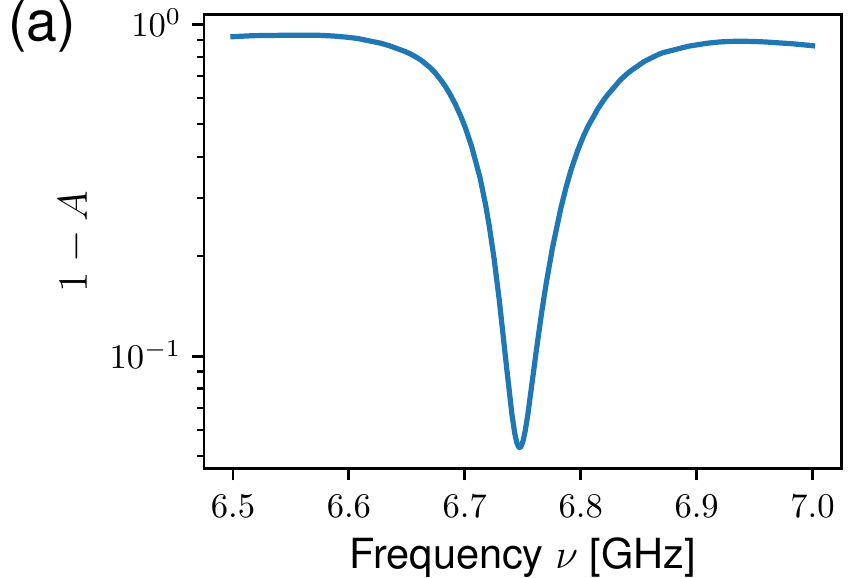}
		\includegraphics[width=.24\linewidth]{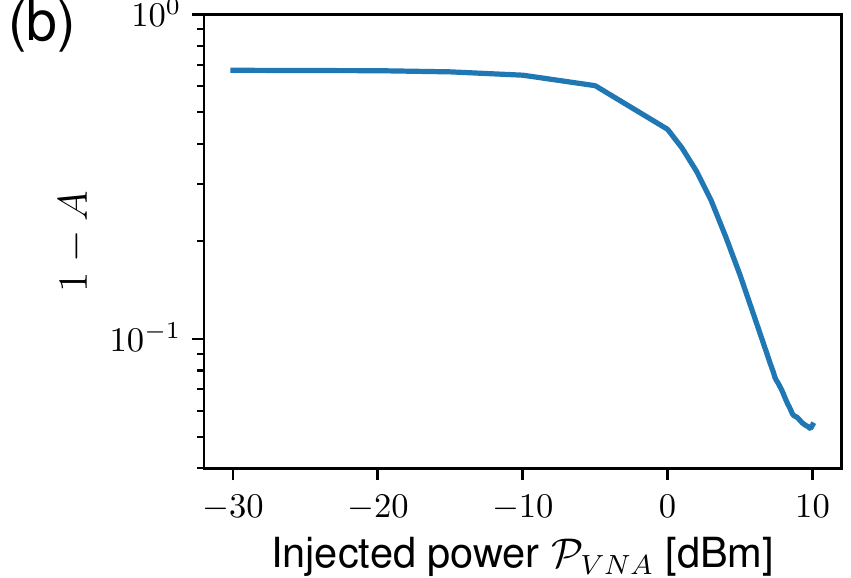}
		\includegraphics[width=.24\linewidth]{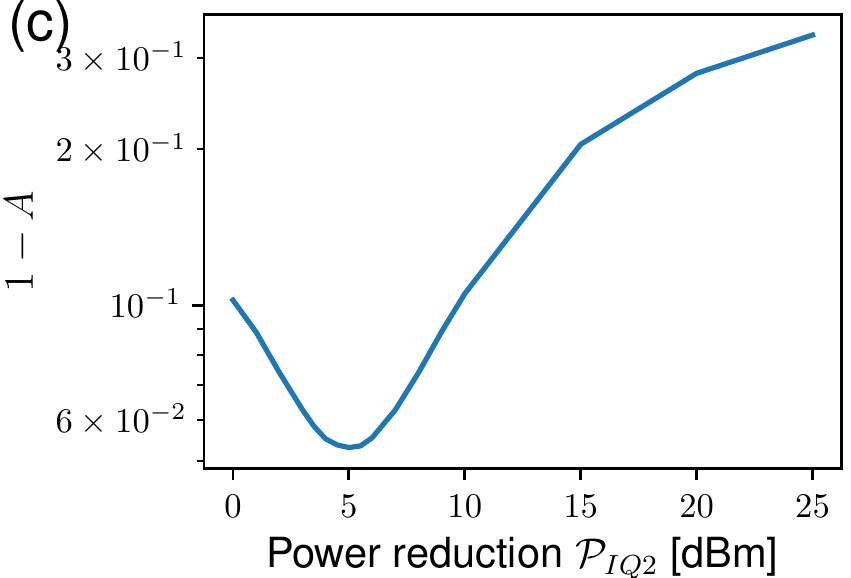}
		\includegraphics[width=.24\linewidth]{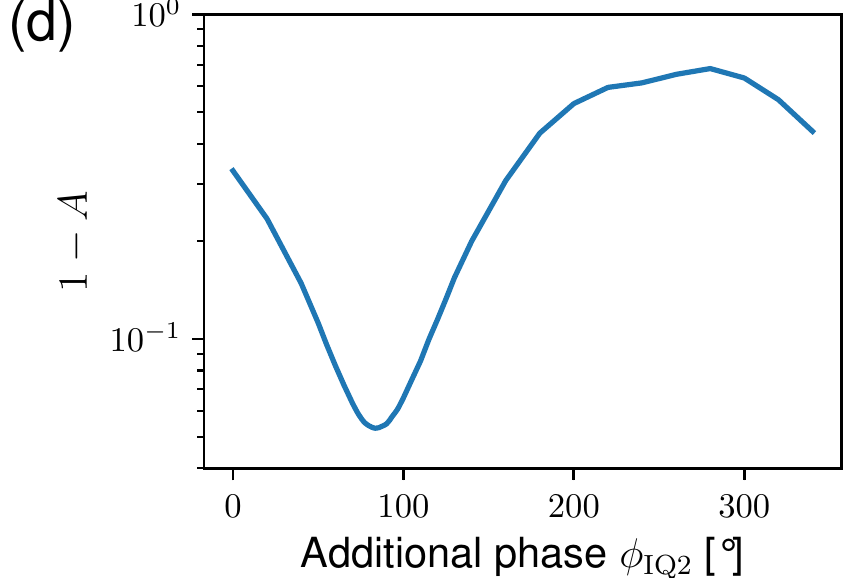}
	}
	\caption{ \label{fig:figs4}
		Deviation from the perfect absorption $1-A$ for the intermediately coupled antennas ($\kappa_0=0.4\nu$) as function of the different 
		experimental parameters frequency $\omega$ (a), injected power $P\uVNA$ by the VNA (b), IQ-modulator intensity reduction ${\cal P}
		\uIQ2$ (c) and phase $\phi\uIQ2$ (d). The other parameters have been fixed to the value where the maximal absorbance (CPA) has 
		been measured, i.e., $P\uNLCPA=9.8$\,dBm, $\omega\uNLCPA\approx 6.747$\,GHz, ${\cal P}\uIQ1=5$\,dB, $\phi\uIQ1=0^\circ$,  
		${\cal P}\uIQ2=5.0$\,dB, and $\phi\uIQ2=83.5^\circ$, respectively.
	}
\end{figure}

In Fig.~\ref{fig:figs3} the deviation from perfect absorbance, i.e., $1-A$ is shown on a logarithmic scale for the weak antenna coupling case 
($\kappa=0.23\nu$) while in Fig.~\ref{fig:figs4} we show our measurements for the intermediate coupling case ($\kappa=0.4\nu$) for different 
slices of the experimental parameters. 
In all cases the other parameters are kept fixed -- taking values corresponding to the minimal absorbance value, i.e., each slice is going through the CPA point. 
For the weak coupling case, we observe a sharp resonance behavior of the absorption whereas for the moderate coupling case the resonance features are smoothed out and become broadened.

\newpage
\section{NLCPA frequencies and constraints for the form of the non-linear losses}

Equation~(7) can be solved by introducing the variable $z=e^{ik}$. 
The corresponding roots take the form
\begin{equation} \label{eq_s:zfunction}
z_{\pm}=\frac{\Omega(|\frac{\nu}{\kappa_1}I_1|^2)\pm \sqrt{\Omega^2(|\frac{\nu}{\kappa_1}I_1|^2)-4(\nu^2-
		\kappa^2)}}{2\nu}\,,
\end{equation}
which after substitution back to Eq.~(7) allows to evaluate the potential NLCPA frequencies $\omega\uNLCPA$ in terms of the intensities of the incident waves. 

The additional requirement that the NLCPA frequencies have to support propagating waves at the leads, i.e., $\omega\uNLCPA \in {\cal R}$, imposes a number of constraints for the form of the lossy nonlinearities. 
Specifically
\begin{equation}
\left(\frac{\Omega_r}{\nu}\right)^2=\left(1-\left(\frac{\nu}{\kappa^2}\Omega_i\right)^2\right) \left(2-\frac{\kappa^2}{\nu^2}\right)^2.
\end{equation}
In particular, if the couplings satisfy the relation $2\nu^2=\kappa^2=|\kappa_1|^2+|\kappa_2|^2$, the nonlinearity can take any purely imaginary form (i.e. $\Omega_r=0$). 
Otherwise, the NLCPA's can occur when the following condition is satisfied 
\begin{equation} \label{eq_s:main_one}
\left(\frac{\Omega_r}{2\nu-\kappa^2/\nu}\right)^2+\left(\frac{\Omega_i}{\kappa^2/\nu}\right)^2=1.
\end{equation}

\end{document}